\documentclass[journal]{IEEEtran}
\ifCLASSINFOpdf
\usepackage{graphicx}
\else
\fi
\usepackage{float}

\usepackage{graphicx}
\usepackage{subcaption}

\usepackage{epstopdf}
\usepackage{color,soul}

\begin{document}
%
\title{ \fontsize{18}{0.5em}\selectfont Noise characterization of an Optical Frequency Comb  using Offline Cross-Correlation}
%
%
%

\author{Ramin Khayatzadeh, Mathieu Collombon, Didier Guyomarc'h, Didier Ferrand, Ga\"etan Hagel, Marie Houssin, Olivier Morizot, Caroline Champenois, and Martina Knoop
\thanks{All authors are with Aix Marseille Universit\'{e}, CNRS, PIIM, UMR 7345, 13397 Marseille, France email: ramin.khayatzadeh@univ-amu.fr, marie.houssin@univ-amu.fr.}}

\maketitle

\begin{abstract}
Using an offline cross-correlation technique, we have analyzed the noise behavior of a new type of optical frequency comb (OFC), which is carrier envelope offset (CEO) free by configuration, due to  difference frequency generation. In order to evaluate the instrument's ultimate  noise floor, the phase and amplitude noise of a stabilized OFC are measured simultaneously  using two analog-to-digital converters. Carrier recovery and phase detection are done by post-processing, eliminating the need for external phase-locked loops and complex calibration techniques. In order to adapt the measurement noise floor and the number of averages used in cross correlation, an adaptive frequency resolution for noise measurement is applied. Phase noise results are in excellent agreement with measurements of the fluctuations of the repetition frequency of the OFC obtained from optical signal.     
\end{abstract}


%
\IEEEpeerreviewmaketitle

\section{Introduction}
%
%
%
%
\IEEEPARstart{O}{ptical} frequency combs (OFCs) find a multitude of applications in different scientific and engineering domains. One of the prime 
applications of OFCs is optical frequency measurement where an OFC, thanks to its large frequency span, behaves as a ruler to measure frequency differences \cite{udem02,pastor04}. Frequency combs that emit at telecommunication wavelengths are especially attractive for microwave-photonics and radio-over-fibre communication systems in order to generate millimetre wave and low-phase-noise carriers which can hugely increase communication data rates \cite{stoehr09,khayatzadeh14}. Recently, offset-free OFCs have been introduced in which the carrier-envelope offset (CEO) frequency, $f_0$, cancels out in a non-linear process of difference frequency generation (DFG) \cite{nakamura15,Puppe16,krauss11}. Ultimate performances on all OFC modes can be reached by locking the OFC to an ultra-stable frequency reference \cite{mcferran05} by means of a single parameter, the repetition frequency  $f_{rep}$.
 Because of the practical importance of OFCs for the mentioned applications, the characterization of noise in OFC lasers has received considerable attention. There are many techniques in both RF and optical domains for measuring the phase and amplitude noise of OFCs. Traditionally, phase noise measurements are made by analogue techniques \cite{Agilent11,Walls88,Howe81}. Recently new approaches based on digital measurements have attracted many attentions. For example, in \cite{Khayatzadeh13}, the authors present an accurate phase-noise measurement technique based on sampling and post-processing which is able to separate amplitude and phase noise measurements. However, when the phase noise of the device under test (DUT) is much lower than the one of the measurement device, the measure is limited by the measurement device noise. This impact can be removed by implementing a cross-correlation technique \cite{Rubiola00,Feldhaus16}. In \cite{grove04}, a real-time direct digital phase noise measurement device based on cross correlation is developed.
 This set-up uses two analog-to-digital converters (ADC) in order to sample signal from the DUT, and then measures phase noise  by using a digital phase detector and applying cross correlation via a FPGA processor.  

In this paper, a  noise detection and cross-correlation technique is presented to measure the amplitude and phase noise of a stabilized OFC using only two ADCs. The described method has been  implemented completely offline. Comparing this technique to the real-time analog techniques, there is no need of phase-locked loops and phase detectors at the inputs since the phase and amplitude detection are performed by post-processing. Other advantages of this technique can be mentioned as follows: the oscillator noise can be compared at different frequencies, amplitude and phase noise can be measured simultaneously and the cost of this cross correlator is largely inferior to commercial devices. The measured results are compared with the direct optical evaluation presented in \cite{Puppe16} and a very good agreement is found.

\section{Set-up scheme and preliminary tests}
\subsection{Configuration of the cross-correlation}

\begin{figure}
\advance\leftskip+0mm
\includegraphics[width=88 mm,height=35 mm]{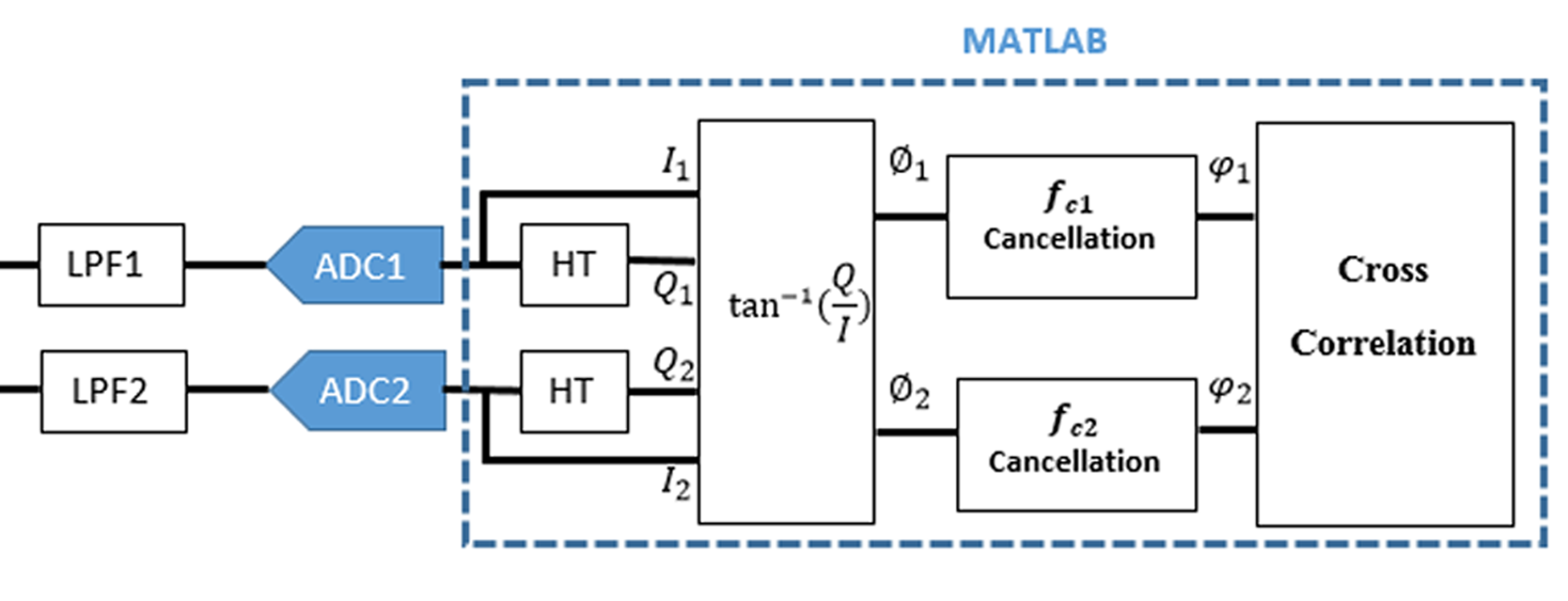}
\caption{Block diagram of offline noise detector base on cross correlation, LPF: Low Pass Filter, ADC: Analog to Digital Converter, HT: Hilbert Transfer.}
\label{fig:1MATLAB_code}
\end{figure}
Figure 1 presents the schematic diagram of the measurement set-up. In this diagram, the output signal of the DUT is divided into two parts and after passing through anti-harmonics and anti-aliasing filters (LPF1 and LPF2), each part is sampled using an ADC. Both ADCs used in this set-up  are independent and they work using different clocks. After sampling, a complete post-processing approach is used to detect the phase noise of the DUT using an I/Q phase detector and cross correlation technique. These processes are presented in the dashed box in Fig. \ref{fig:1MATLAB_code}. In the first step, the quadrature component of the sampled signal is determined using the Hilbert transfer function (HT). Then, knowing the in-phase and quadrature components, the phase of the sampled signal ($\phi_{1}$ and $\phi_{2}$) can be determined by using unwrapped arctangent function. In the next step, the carrier frequency is removed from the detected phase whose general expression is
\begin{equation}
\phi (t)=2 \pi f_{c} t+\varphi(t)
\label{Eq:1}
\end{equation}

where $f_{c}$ is the carrier frequency and $\varphi(t)$ is the detected phase noise.
 
 We need to mention that since the detected phase using the I/Q detector in the previous step is unwrapped, its value contains the angular frequency multiplied by time values (see Eq. \ref{Eq:1}). Thus, considering that the carrier frequency is much higher than the frequency fluctuations, one can find the exact value of the carrier frequency utilizing a linear polynomial curve fitting technique.

After determining $f_{c}$ and removing the term $2 \pi f_{c} t$, one can extract the phase noise value, $\varphi(t)$, from Eq.(\ref{Eq:1}).  The cross correlation technique is applied in the next step  to reduce the noise floor of the measurement system.

Additionally, the amplitude noise $\epsilon(t)$ can be measured using an envelope detector based on in-phase and quadrature components as follow:

\begin{equation}
\epsilon (t)=\frac{\sqrt{I^2+Q^2}-A}{A}
\label{Eq:2}
\end{equation}

where I and Q are the in-phase and quadrature components and A is the mean amplitude of the signal.

\subsection{Simulation and calibration}

\begin{figure}
\includegraphics[width=84 mm]{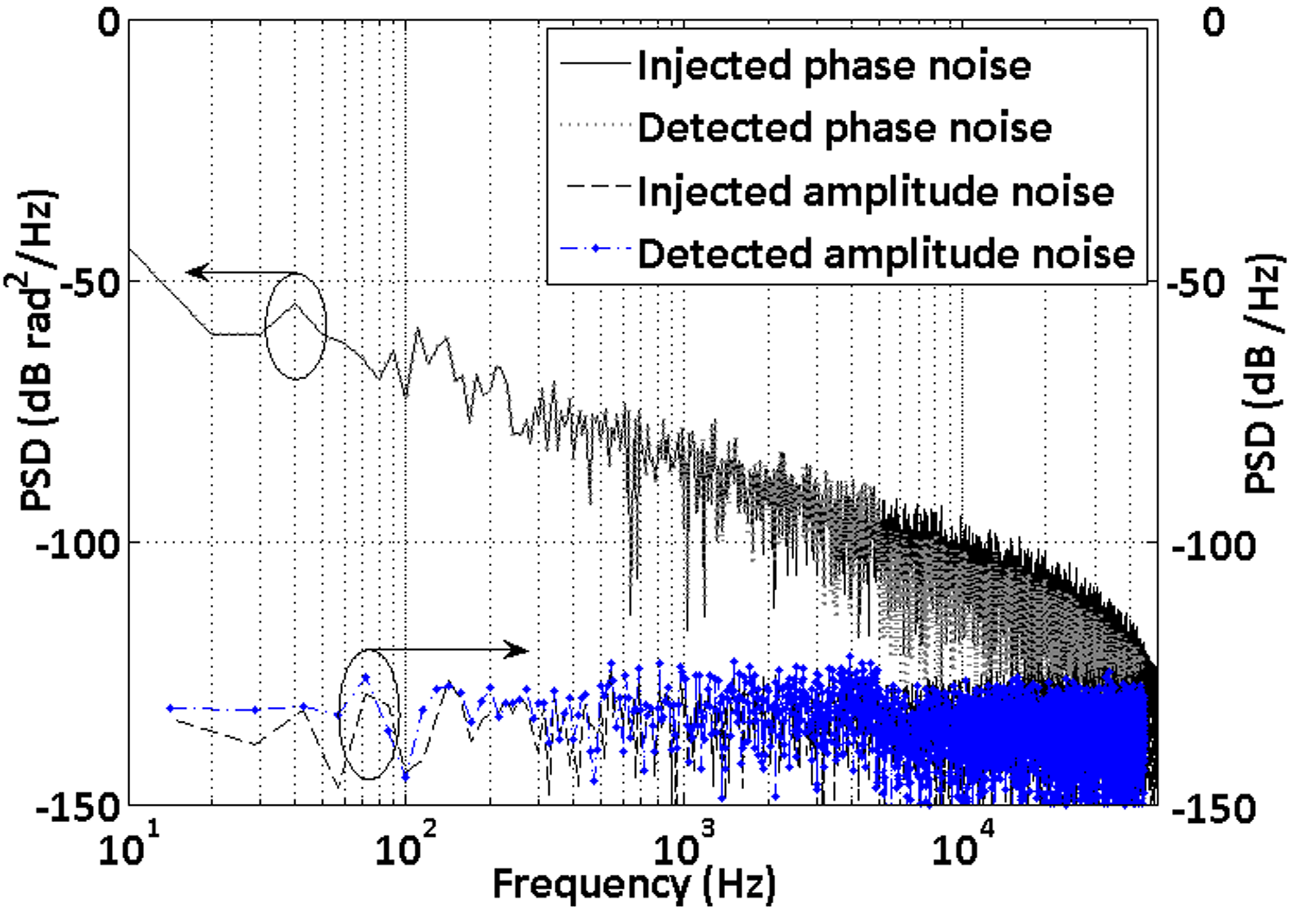}
\caption{PSD of the injected and detected phase and amplitude noise.}
\label{fig:2Simulation}
\end{figure}

Prior to measurements on the OFC, a simulation is performed in order to verify the accuracy of phase and amplitude detectors used in the  cross correlation technique. To check the phase noise detector, a sine function signal is corrupted with phase and amplitude noise and by using the phase and amplitude detectors the power spectral density (PSD) of the detected noise and the injected noise are compared. The phase noise is generated by integrating a white noise signal giving a slope of the PSD of phase noise at -20 dB per decade which presents the impact of $1/f$ noise. The amplitude noise is considered to be white Gaussian noise. Figure \ref{fig:2Simulation} illustrates the PSD of the injected phase noise (black bold curve) and the detected phase noise (gray dashed curve) and the PSD of the injected (black dash curve) and detected (dashed dot curve) amplitude noise. As can be seen, the noise detected by the described measurement set-up and the injected noise are identical, which confirms the accuracy of the phase and amplitude detectors. This simulation permits us to calibrate the phase and amplitude noise detectors.

\section{Experimental set-up}

\begin{figure}
\includegraphics[width=86 mm]{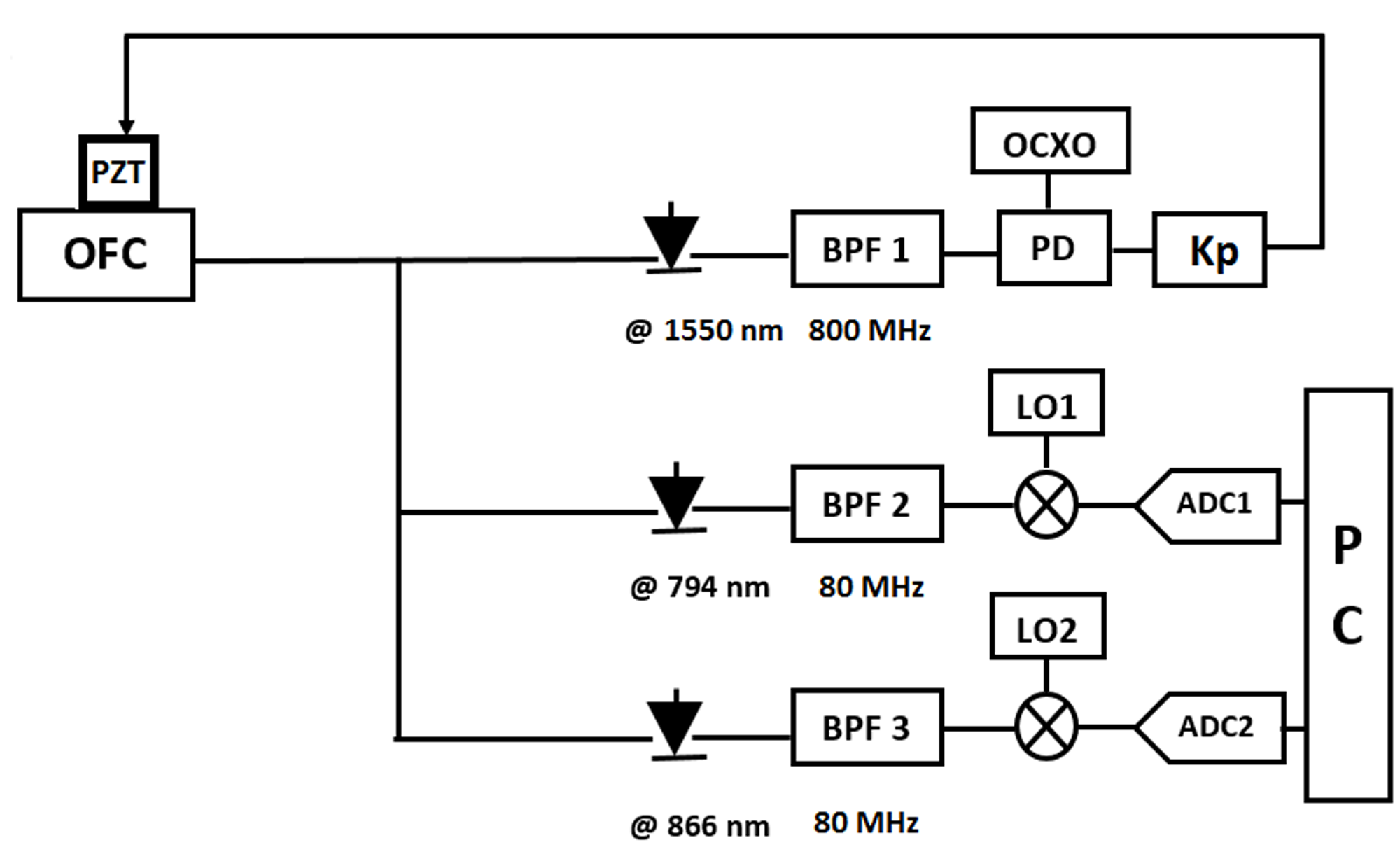}
\caption{Experimental set-up of the noise measurement system, OFC: Optical Frequency Comb, BPF: Band Pass Filter, PD: Phase Detector, ADC: Analog to Digital Converter, OCXO: Oven-Controlled Crystal Oscillator, PZT: Piezoelectric Transducer, LO: Local Oscillator.}
\label{fig:3ExperimentDiagram}
\end{figure}

The block diagram of the experimental set-up of the  measurement system is presented in Fig. \ref{fig:3ExperimentDiagram}. Passive mode-locking \cite{mollenauer84} of an Erbium fiber laser at 1550 nm generates femto-second pulses which are the basis of the OFC. After spectral broadening in a photonic crystal fiber \cite{topticaDFC} and DFG \cite{nakamura15,Puppe16,krauss11}, which both preserve the phase relationship between all modes, a super-continuum light spanning from 1450 nm to 1625 nm is obtained. It is composed by equally spaced modes separated by the repetition frequency $f_{rep}$ without offset frequency,   mode $N$ has a frequency $f_N = N\times f_{rep}$. Other phase-coherent processes, allow to access the wavelengths used in our set-up designed for precision spectroscopy of trapped Ca$^+$-ions \cite{champenois07}. For long-term stability,   the tenth harmonics of the repetition frequency $10 f_{rep}$ at 800 MHz, filtered by a band-pass filter (BPF1), is phase-locked to a low-noise crystal oscillator referenced to GPS (Global positioning system). Corrections of the servo system are applied to the piezo-electric element of the OFC laser after applying proportional control gain ($K_p$).

\begin{figure}
\advance\leftskip+0mm
\includegraphics[width=86 mm]{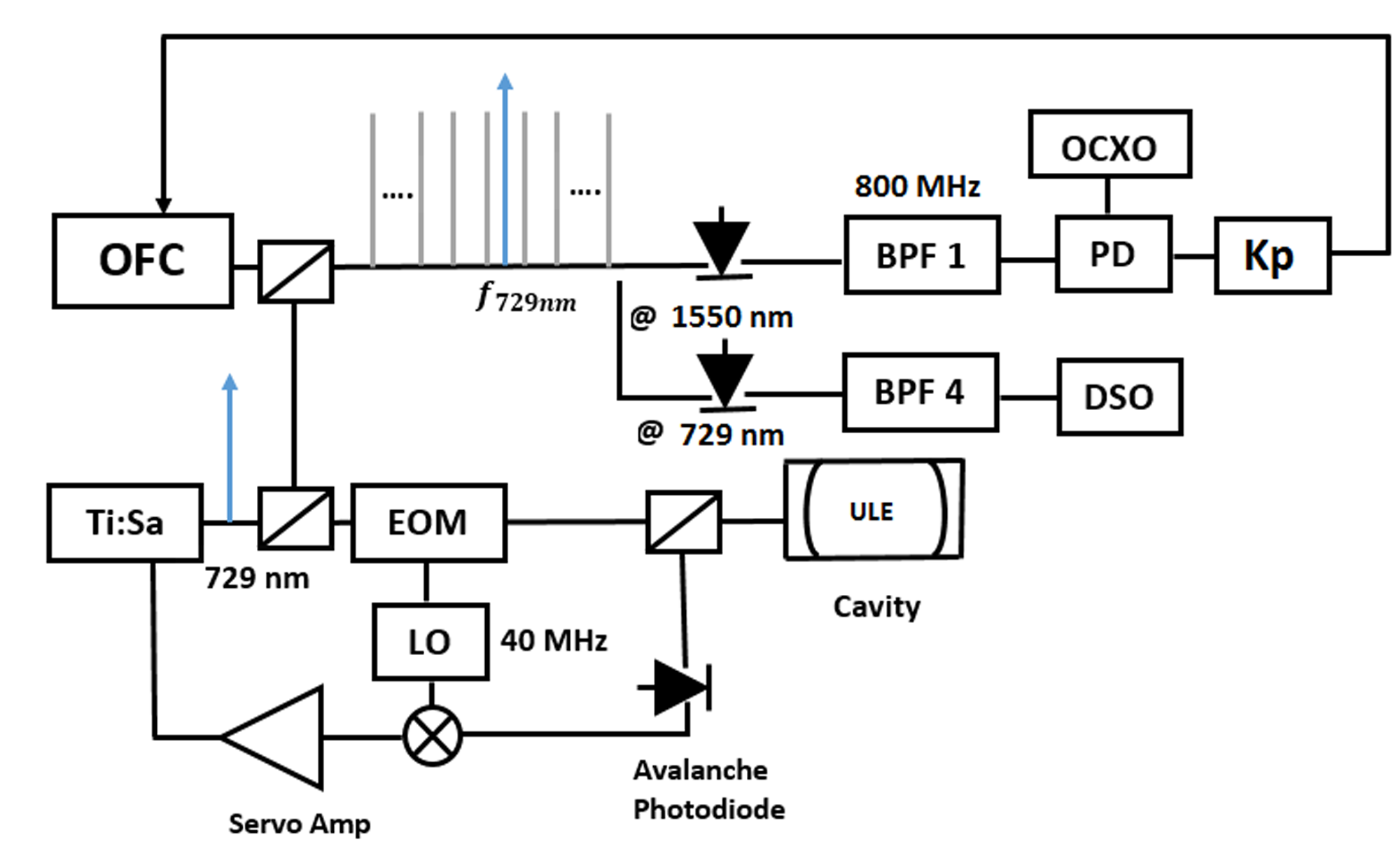}
\caption{Experimental set-up of the optical noise measurement system, OFC: Optical Frequency Comb, BPF: Band Pass Filter, PD: Phase Detector, DSO: Digital Storage Oscilloscope, OCXO: Oven-Controlled Crystal Oscillator, EOM: Electro Optic Modulator, LO: Local Oscillator, ULE: Ultra Low Expansion.}
\label{fig:3_2ExperimentDiagram}
\end{figure}

To investigate $f_{rep}$ noise  in a  cross-correlation method, one needs to detect $f_{rep}$ with two fully uncorrelated experimental benches. We use two photodiodes at 794 nm and 866 nm, to detect the OFC signal via 3-meter long single-mode, polarization maintaining fibers which are shorter than the coherence length of the signal. Each mode beats with all its neighbors, and therefore each photodiode signal is composed of a fundamental beat at 80 MHz and harmonics. After band-pass filtering (BPF2 and BPF3 in Fig.\ref{fig:3ExperimentDiagram}) a beat note is obtained  at $f_{rep}$=80 MHz exhibiting a 40~dB signal to noise ratio. 
In the next step, the two signals are frequency down-converted to lower frequencies (250 Hz) in order to be sampled by two analog to digital converters (ADC). The local oscillators (LO1 and LO2) are two independent signal generators (HP 8657A and Marconi 2022C) and the ADCs are two National Instrument acquisition cards (16 bit, 2 Msps) driven by two different clock signals. After signal acquisition, a complete off-line post-processing cross-correlation analysis is applied to eliminate all non-common noises (PD, LO, down-conversion, ADC noises) which reduces the noise floor of the measurement and allows to determine the phase noise of the repetition frequency $f_{rep}$. 

\begin{figure}
\advance\leftskip+0mm
\includegraphics[width=88 mm,height=60 mm]{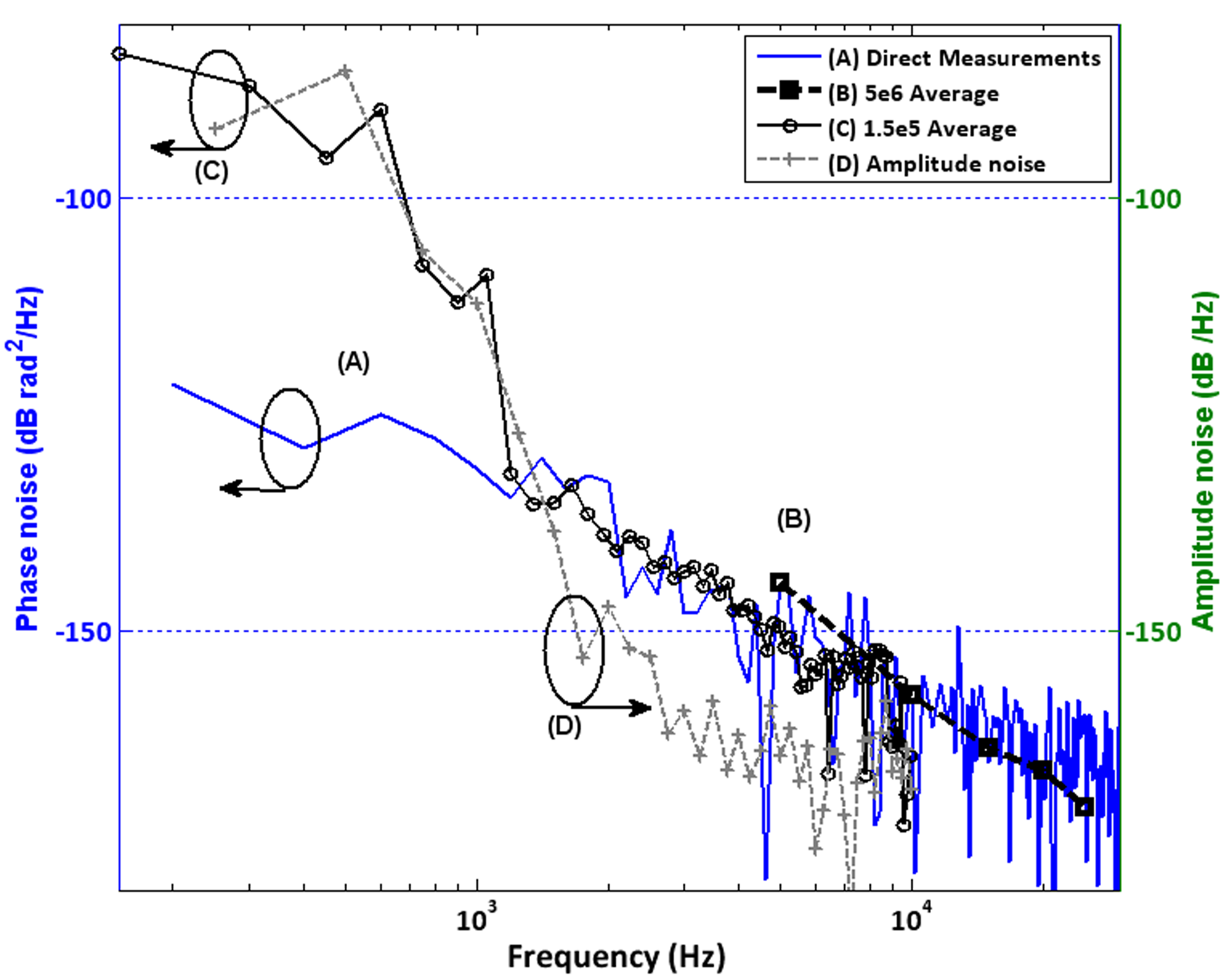}
\caption{Power spectral density of the phase noise for (A) Optical measurement (B) Cross correlation 5.10$^6$ average (C) Cross correlation 2.10$^5$ average and (D) the amplitude noise results 2.10$^5$ average.}
\label{fig:4NoiseResults}
\end{figure}

In order to verify the accuracy of the measurement results, the PSD of the phase noise of the repetition frequency is deduced from a phase noise measurement directly in the optical domain. The block diagram of the set-up is depicted in Fig. \ref{fig:3_2ExperimentDiagram} and the results are compared with those of the cross-correlation technique (See Fig. \ref{fig:4NoiseResults}). The optical measurement is performed utilizing an ultra-stable frequency Titanium-sapphire (Ti:Sa) laser as a reference. The frequency of the laser is stabilized using an ultra low expansion (ULE) cavity and applying Pound-Drever-Hall technique. This laser has a linewidth  of 2 Hz (measured at resolution frequency of 1 Hz) and a frequency stability of better than $2.10^{-14}$ at 1 second. In the optical measurement technique, we record the beat note signal between the reference laser signal at 729~nm and the nearest optical mode of the CEO free OFC at $f_{729nm}$ ($f_{729nm}=N_{729nm} \times f_{rep}$, $N_{729nm}=5137500$). This beat-note is filtered using BPF4 (around 48~MHz) and then  sampled using a digital storage oscilloscope (DSO) in order to measure the PSD of its phase noise. Since its phase noise is $N_{729nm}$ times higher than that of the repetition frequency and so its PSD is $N_{729nm}^2$ times the PSD of the repetition frequency, as it is shown and proved in \cite{Puppe16} for the same type of OFC, the PSD of the beat note can be measured directly. For comparison to the cross-correlation data, this measured PSD is then mathematically divided by $N_{729nm}^2$.

\section{Results and discussion}

The PSD of the phase and amplitude noise of the repetition frequency $f_{rep}$ of the OFC  under test is presented in Fig. \ref{fig:4NoiseResults}. Curve (A) shows the phase noise PSD measured by the direct optical measurement technique. The curves (C) and (B) present the PSD of the measured phase noise for two different frequency resolutions and different number of averages used in cross correlation analysis. For the curve (C), since the phase-noise values are higher for the offset frequencies close to the carrier (below 10 kHz), a lower number of averages ($2 \times 10^5$) is sufficient to measure the real phase-noise value, the frequency resolution is then 150 Hz. In contrary, since the phase-noise values for the offset frequencies higher than 10 kHz are much lower than those of frequencies closer to zero Hz, the number of averages needs to be increased to $5 \times 10^6$ in order to reach the phase noise values of repetition frequency. The resolution is then 5 kHz. The problem that arises when using a higher number of averages is that the frequency resolution will be less due to the finite number of samples. It appears that an adaptation of the number of averages as a function of the noise frequency range is essential. As can be seen, the PSDs obtained from the cross correlation technique and the indirect measurement result are identical for offset frequencies higher than 1 kHz. Phase noise value of -157 dB $rad^2$/Hz is measured at 10 kHz. The differences below 1 kHz are a consequence of harmonics of the carrier signal (250 Hz) at 500 Hz, 750 Hz and 1 kHz in the cross correlation measurement chain  which originate from the frequency down conversion and ADC sampling process. Curve (D) presents the PSD of the amplitude noise measured using the cross correlation technique. The impact of the harmonics can be seen on this curve, too. The amplitude noise values for the frequencies from 1 kHz to 10 kHz are much lower than the phase noise values (approximately -170 dB/Hz at 10 kHz).

\begin{figure}
\advance\leftskip+0mm
\includegraphics[width=85 mm,height=60 mm]{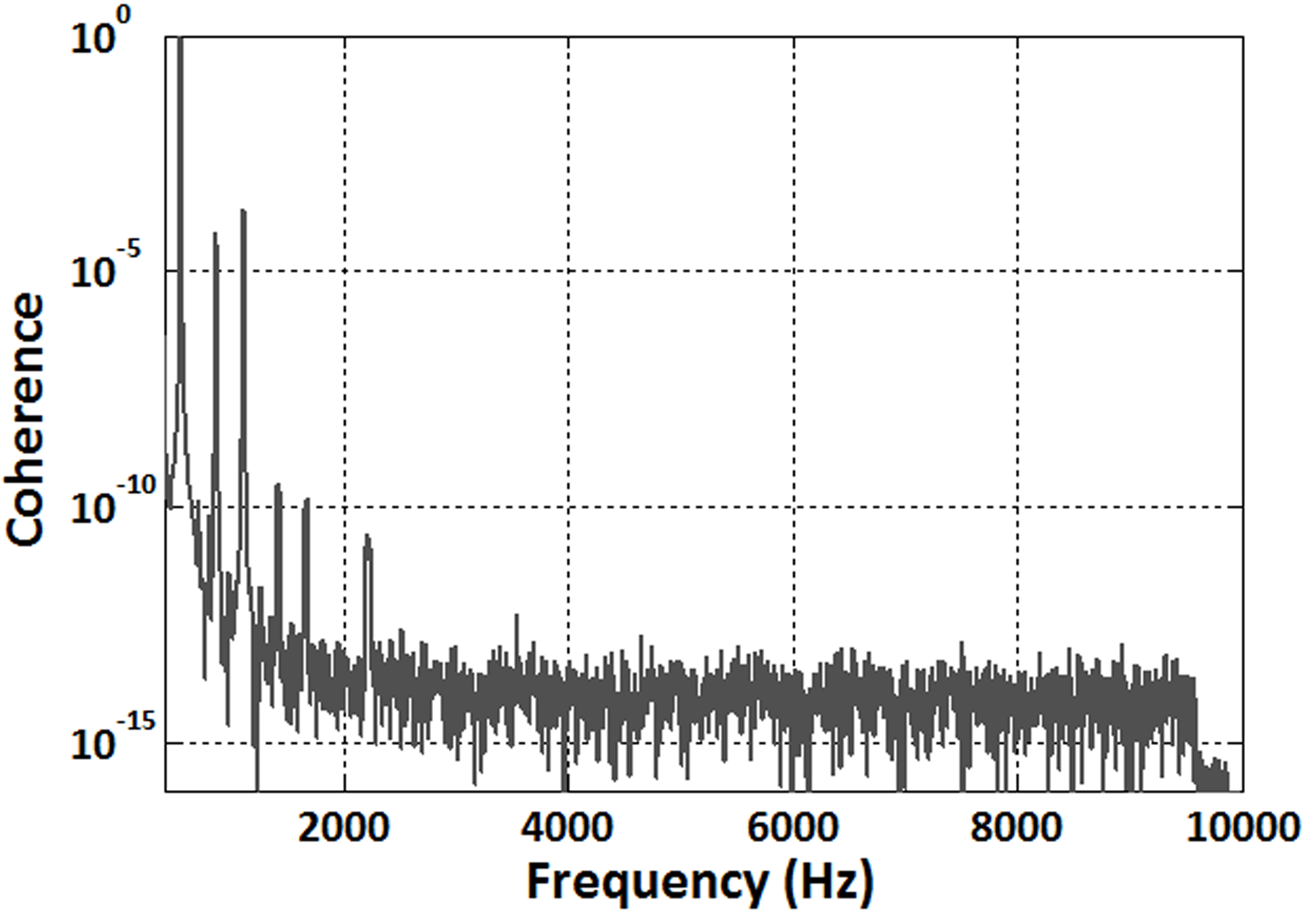}
\caption{Coherence function between amplitude and phase noise measurements.}
\label{fig:6AMPMRcoherence}
\end{figure}

In order to illustrate the correlation between the amplitude and phase noise a coherence function is calculated using the following equation \cite{Tsuchida98}:

 \begin{equation}
\gamma ^2 (f)=\frac{S_{\epsilon\phi}^*(f) S_{\epsilon\phi}(f)}{S_{\epsilon}(f)S_{\phi}(f)} 
\end{equation} 

where $\gamma^2$ is a normalized parameter such that the values 1 and 0 mean perfect and no correlation, respectively. In this equation, $S_{\epsilon}(f)$, $S_{\phi}(f)$, and $S_{\epsilon\phi}(f)$, are the PSDs of the amplitude noise, phase noise and the cross spectral density between the amplitude and phase noise, respectively.

The coherence function result is presented in Fig. \ref{fig:6AMPMRcoherence}. As can be seen, there is no significant coherence between the phase and amplitude noise for  frequencies higher than 1 kHz. This lack of coherence confirms that the amplitude noise has no impact on the phase noise due to its very low level. On the contrary, relatively large peaks can be found at harmonics of 250 Hz. The strong coherence at these frequencies confirms that these harmonics on amplitude and phase noise have the same origin, as has been observed on Fig. \ref{fig:4NoiseResults} for both curves (C) and (D). Since these harmonics are missing in the optical measurement results, there is evidence that they are artifacts due to the frequency down-conversion and sampling process in the cross-correlation chain.

\section{Conclusion}
In this paper, a complete offline cross-correlation phase and amplitude noise measurement technique is presented to characterize a CEO free optical frequency comb. The output signal of the device under test is sampled by standard data acquisition cards and all the measurement processes such as phase detection, amplitude detection, and cross correlation are performed by post processing. The cost and complexity of the proposed technique are lighter compared to commercial digital and analog techniques. The accuracy of the measurement results is verified using an optical measurement technique and a very good agreement between the results is observed. Cross correlation analysis is also applied  to observe the coherence between the amplitude and phase noise measurements, showing a high degree of independence for frequencies higher than 1 kHz.

\section*{Acknowledgments}
This work has been carried out thanks to the support of the A*MIDEX project (n$^\circ$ ANR-11-IDEX-0001-02), EquipEx Refimeve+ (n$^\circ$ ANR-11-EQPX-0039), and Labex First-TF (n$^\circ$ ANR-10-LABX-48-01), all funded by the "Investissements d'Avenir" French Government program, managed by the French National Research Agency (ANR).


%




\bibliographystyle{IEEEtran}

\begin{thebibliography}{10}
\providecommand{\url}[1]{#1}
\csname url@samestyle\endcsname
\providecommand{\newblock}{\relax}
\providecommand{\bibinfo}[2]{#2}
\providecommand{\BIBentrySTDinterwordspacing}{\spaceskip=0pt\relax}
\providecommand{\BIBentryALTinterwordstretchfactor}{4}
\providecommand{\BIBentryALTinterwordspacing}{\spaceskip=\fontdimen2\font plus
\BIBentryALTinterwordstretchfactor\fontdimen3\font minus
  \fontdimen4\font\relax}
\providecommand{\BIBforeignlanguage}[2]{{%
\expandafter\ifx\csname l@#1\endcsname\relax
\typeout{** WARNING: IEEEtran.bst: No hyphenation pattern has been}%
\typeout{** loaded for the language `#1'. Using the pattern for}%
\typeout{** the default language instead.}%
\else
\language=\csname l@#1\endcsname
\fi
#2}}
\providecommand{\BIBdecl}{\relax}
\BIBdecl

\bibitem{udem02}
T.~Udem, R.~Holzwarth, and T.~H{\"a}nsch, ``Optical frequency metrology,''
  \emph{Nature}, vol. 416, pp. 233--237, 2002.

\bibitem{pastor04}
P.~C. Pastor, G.~Giusfredi, P.~D. Natale, G.~Hagel, C.~de~Mauro, and
  M.~Inguscio, ``Absolute frequency measurements of the
  ${2}^{3}{S}_{1}\ensuremath{\rightarrow}{2}^{3}{P}_{0,1,2}$ atomic helium
  transitions around 1083 nm,'' \emph{Phys. Rev. Lett.}, vol.~92, p. 023001,
  Jan 2004.

\bibitem{stoehr09}
A.~St\"ohr, A.~Akrout, R.~Bu\ss, B.~Charbonnier, F.~van Dijk, A.~Enard,
  S.~Fedderwitz, D.~J\"ager, M.~Huchard, F.~Lecoche, J.~Marti, R.~Sambaraju,
  A.~Steffan, A.~Umbach, and M.~Wei\ss, ``60 {GH}z radio-over-fiber
  technologies for broadband wireless services,'' \emph{J. Opt. Netw.}, vol.~8,
  pp. 471--487, 2009.

\bibitem{khayatzadeh14}
R.~Khayatzadeh, J.~Poette, and B.~Cabon, ``Impact of phase noise in 60-{GH}z
  radio-over-fiber communication system based on passively mode-locked laser,''
  \emph{IEEE J. Lightw. Technol.}, vol.~32, pp. 3529--3535, 2014.

\bibitem{nakamura15}
T.~Nakamura, I.~Ito, and Y.~Kobayashi, ``Offset-free broadband {Y}b:fiber
  optical frequency comb for optical clocks,'' \emph{Opt. Express}, vol.~23,
  pp. 19\,376--19\,381, 1990.

\bibitem{Puppe16}
T.~Puppe, A.~Sell, R.~Kliese, N.~Hoghooghi, A.~Zach, and W.~Kaenders,
  ``Characterization of a {DFG} comb showing quadratic scaling of the phase
  noise with frequency,'' \emph{Opt. Lett.}, vol.~41, no.~8, pp. 1877--1880,
  Apr 2016.

\bibitem{krauss11}
\BIBentryALTinterwordspacing
G.~Krauss, D.~Fehrenbacher, D.~Brida, C.~Riek, A.~Sell, R.~Huber, and
  A.~Leitenstorfer, ``All-passive phase locking of a compact {E}r: fiber laser
  system,'' \emph{Optics letters}, vol.~36, no.~4, pp. 540--542, 2011.
  [Online]. Available:
  \url{http://www.opticsinfobase.org/abstract.cfm?uri=ol-36-4-540}
\BIBentrySTDinterwordspacing

\bibitem{mcferran05}
J.~J. {McFerran}, E.~N. Ivanov, A.~Bartels, G.~Wilpers, C.~W. Oates, S.~A.
  Diddams, and L.~Hollberg, ``Low-noise synthesis of microwave signals from an
  optical source,'' \emph{Electronics Letters}, vol.~41, no.~11, p.~1, 2005.

\bibitem{Agilent11}
``Agilent’s phase noise measurement solutions,,'' \emph{Agilent Technol.},
  2011.

\bibitem{Walls88}
F.~L. Walls, A.~J. Clements, C.~M. Felton, M.~A. Lombardi, and M.~D. Vanek,
  ``Extending the range and accuracy of phase noise measurements,'' in
  \emph{Proc. 42nd Ann. Freq. Control. Symp.}, 1988.

\bibitem{Howe81}
D.~H. Howe, D.~Allan, and J.~A. Barnes, ``Properties of signal source and
  measurement methods,'' in \emph{Proc. 35th Ann. Symp. Freq. Control}, May
  1981.

\bibitem{Khayatzadeh13}
R.~Khayatzadeh, H.~Rzaigui, J.~Poette, and B.~Cabon, ``Accurate millimeter-wave
  laser phase noise measurement technique,'' \emph{IEEE Photon. Technol.
  Lett.}, vol.~25, pp. 1218--1221, 2013.

\bibitem{Rubiola00}
E.~Rubiola and V.~Giordano, ``Correlation-based phase noise measurements,''
  \emph{Rev. Sci. Instrum.}, vol.~71, p. 3085–3091, August 2000.

\bibitem{Feldhaus16}
G.~Feldhaus and A.~Roth, ``A 1{MHz} to 50 {GHz} direct down-conversion phase
  noise analyzer with cross-correlation,'' in \emph{2016 European Frequency and
  Time Forum (EFTF)}, April 2016, pp. 1--4.

\bibitem{grove04}
J.~Grove, J.~Hein, J.~Retta, P.~Schweiger, W.~Solbrig, and S.~Stein,
  ``Direct-digital phase-noise measurement,'' in \emph{Frequency Control
  Symposium and Exposition, 2004. Proceedings of the 2004 IEEE International},
  Aug 2004, pp. 287--291.

\bibitem{mollenauer84}
L.~F. Mollenauer and R.~H. Stolen, ``The soliton laser,'' \emph{Opt. Lett.},
  vol.~9, pp. 13--15, 1984. 

\bibitem{topticaDFC}
\BIBentryALTinterwordspacing
{Toptica GmbH}. (2015) Difference frequency comb. [Online]. Available:
  \url{http://www.toptica.com/products/frequency-combs/dfc-core.html}
\BIBentrySTDinterwordspacing

\bibitem{champenois07}
C.~Champenois, G.~Hagel, M.~Houssin, M.~Knoop, C.~Zumsteg, and F.~Vedel,
  ``Terahertz frequency standard based on three-photon coherent population
  trapping,'' \emph{Phys. Rev. Lett.}, vol.~99, p. 013001, 2007.

\bibitem{Tsuchida98}
H.~Tsuchida, ``Correlation between amplitude and phase noise in a mode-locked
  {C}r:{L}i{SAF} laser,'' \emph{Opt. Lett.}, vol.~23, no.~21, pp. 1686--1688,
  1998.

\end{thebibliography}
\end{document}